\documentstyle[psfig, epsfig, 12pt]{article}
\def\del{{\partial}}

\newfont{\Bbb}{msbm10 scaled 1200}     
\newcommand{\mathbb}[1]{\mbox{\Bbb #1}}

\def\lbldef#1#2{\expandafter\gdef\csname #1\endcsname {#2}}
\def\eqn#1#2{\lbldef{#1}{(\ref{#1})}%
\begin{equation} #2 \label{#1} \end{equation}}

\def\href#1#2{#2}

\newcommand{\beq}{\begin{equation}}
\newcommand{\eeq}{\end{equation}}
\newcommand{\ber}{\begin{eqnarray}}
\newcommand{\eer}{\end{eqnarray}}
\newcommand{\beqar}{\begin{eqnarray}}

\newcommand{\eeqar}{\end{eqnarray}}

\newcommand{\ba}{\begin{eqnarray}}
\newcommand{\ea}{\end{eqnarray}}

\newcommand{\dsl}
  {\kern.06em\hbox{\raise.15ex\hbox{$/$}\kern-.56em\hbox{$\partial$}}}

\newcommand{\eeqarr}{\end{eqnarray}}
\newcommand{\ZZ}{{\rm \kern 0.275em Z \kern -0.92em Z}\;}

\let\a=\alpha \let\b=\beta \let\g=\gamma \let\d=\delta \let\e=\epsilon
    \let\k=\kappa
 \let\m=\mu \let\n=\nu   \let\r=\rho
\let\s=\sigma \let\t=\tau  \let\f=\phi \let\c=\chi

\def\nn{\nonumber} \def\bd{\begin{document}} \def\ed{\end{document}}
\def\ds{\documentstyle} \let\fr=\frac \let\bl=\bigl \let\br=\bigr
\let\Br=\Bigr \let\Bl=\Bigl 
\let\bm=\bibitem
\let\na=\nabla
\let\pa=\partial \let\ov=\overline 
\newcommand{\be}{\begin{equation}} 
\newcommand{\ee}{\end{equation}} 
\def\ft#1#2{{\textstyle{{\scriptstyle #1}\over {\scriptstyle #2}}}}
\def\fft#1#2{{#1 \over #2}}
\def\vp{\varphi}
\def\sst#1{{\scriptscriptstyle #1}}
\def\oneone{\rlap 1\mkern4mu{\rm l}}
\def\td{\tilde}
\def\wtd{\widetilde}
\def\ie{\rm i.e.\ }
\def\dalemb#1#2{{\vbox{\hrule height .#2pt
        \hbox{\vrule width.#2pt height#1pt \kern#1pt
                \vrule width.#2pt}
        \hrule height.#2pt}}}
\def\square{\mathord{\dalemb{6.8}{7}\hbox{\hskip1pt}}}
\def\wtd{\widetilde}
\def\R{\rlap{\rm I}\mkern3mu{\rm R}}
\def\im{{\rm i}}
\def\tilg{\tilde{g}}
\def\tilF{\tilde{F}}
\def\tilA{\tilde{A}}
\def\varf{\varphi}
\def\tilf{\tilde{\phi}}
\def\tilh{\tilde{h}}
\def\rme{{\rm e}}
\begin{document}
\baselineskip=15.5pt
\pagestyle{plain}
\setcounter{page}{1}
\begin{titlepage}

\leftline{\tt hep-th/0103080}

\vskip -1.cm

\rightline{\small{\tt HUTP-01/A012}}
\rightline{\small{\tt CTP-MIT-3094}}
\rightline{\small{\tt CTP TAMU-09/01}}

\begin{center}

\vskip 0.5 cm

{\LARGE Supergravity duals of gauge theories from $F(4)$ gauged supergravity in six dimensions}

\vskip 1.cm

{\large Carlos N\'u\~nez$^a$, I.Y. Park$^b\dagger$, Martin Schvellinger$^c$, Tuan A. Tran$^b \ddagger$}

\vskip 0.6cm

$^a${\it Lyman Laboratory of Physics, \\
Harvard University, \\
Cambridge, Massachusetts 02138, USA} \\
E-mail: {\tt nunez@lorentz.harvard.edu}

\vskip 0.5cm

$^b${\it Center for Theoretical Physics, \\ 
Texas A$\&$M University \\
College Station, Texas 77843, USA} \\
$\dagger$E-mail: {\tt ipark@rainbow.physics.tamu.edu} \\
$\ddagger$E-mail: {\tt tuan@rainbow.physics.tamu.edu}

\vskip 0.5cm

$^c${\it Center for Theoretical Physics, \\
Laboratory for Nuclear Science and Department of Physics, \\
Massachusetts Institute of Technology, \\
Cambridge, Massachusetts 02139, USA} \\
E-mail: {\tt martin@ctpbeaker.mit.edu}

\vspace{1cm}

{\bf Abstract}

\end{center}

\noindent
  
We construct supergravity solutions dual to twisted field theories
that are the worldvolume theories of D4-branes wrapped on 
2, 3-cycles, and NS-fivebranes on 2-cycles.
We first obtain the solutions for the Romans' six-dimensional
gauged supergravity theories and then up-lift them to 
ten dimensions. In particular, we find solutions for field configurations with
either non-Abelian fields or $B$-fields being excited. One of these solutions, 
in the massless case, is up-lifted to the massless type IIA string theory. 
This is the first example of such a kind. The cases studied
provide new examples of the AdS/CFT duality involving twisted field theories.

\end{titlepage}

\newpage


\baselineskip=15.5pt

\section{Introduction}

The large $N$-limit of superconformal field theories have dual descriptions 
in terms of weakly coupled supergravities
\cite{Maldacena:1998re,Gubser:1998bc,Witten:1998qj}. Generalizations
to non-conformal cases were also studied \cite{Itzhaki:1998dd}. 
There is an extensive list of references regarding the flows of ${\cal{N}}=4$
super Yang-Mills theory under the perturbations of either masses or {\it vev}'s
for bilinear operators \cite{Pilch:2000fu,Freedman:1999gp}. With the twisted gauge field theories
\cite{Bershadsky:1996qy} available, it is possible to find new examples 
of AdS/CFT where only a part of the maximal supersymmetries is preserved
\cite{Maldacena:2000mw,Maldacena:2001yy,Acharya:2000mu,Gauntlett:2000ng,Nieder:2000kc,Edelstein:2001pu}. 
This allows us to investigate supergravity
duals of partially twisted gauge field theories. Therefore, wrapped fivebranes and
D3-branes on holomorphic curves were studied in
\cite{Maldacena:2000mw,Maldacena:2001yy}. Similarly, fivebranes \cite{Acharya:2000mu} 
and D3-branes \cite{Nieder:2000kc} wrapped on associative three-cycles were considered. 
These studies have been extended to M-fivebranes wrapping
K{\"a}hler 4-cycles, special Lagrangian 3, 4 and 5-cycles, 
co-associative 4-cycles and Cayley 4-cycles \cite{Gauntlett:2000ng}.
In this paper, we concentrate on the D4-D8 system, which was studied in
depth from the gauge 
theory side in ref. \cite{Intriligator:1997pq} and from the gravity/CFT viewpoint in refs. 
\cite{Brandhuber:1999np,Ferrara:1998gv}. We will wrap D4-branes of a D4-D8 system on two and three-cycles. 
We will also study solutions corresponding to smeared NS-fivebranes wrapped 
on two-cycles. We first find solutions of six-dimensional Romans' gauged
supergravity \cite{Romans:1986tw} and up-lift them to massive type
IIA string theory. Other constructions in massive type IIA supergravity  can be found in
ref. \cite{Massar:1999sb}, where some intersecting-brane
configurations of D0-branes, D8-branes and a string were studied. 
Then, we analyze their dual twisted 
field theories and study the flows from $AdS_6$-type regions 
to the $AdS_{6-p} \times \Sigma_p$ solutions, 
where $\Sigma_p$ is the $p$-dimensional curved manifold. In the present case we
consider $p=2$ and 3.

The Romans' theories we are interested in are the $SU(2)$ gauged ${\cal N} = 4$ supergravities
in six dimensions. There are four different theories depending upon the values
of the $SU(2)$ coupling constant $g$ and the mass parameter $m$ of
the two-form field $B$. One of these theories has two anti-de Sitter vacua, one of them breaking all the
supersymmetry and the other one preserving all of it. The latter theory, with a ground state with
unbroken $F(4)$ supersymmetry, is the six-dimensional supergravity predicted in
\cite{Nahm:1978tg,DeWitt:1982wm}. In this paper, we will consider this case.

We start out to find solutions of six-dimensional gauged supergravity
with vanishing $B$-fields, using a hyperbolic metric ansatz of
the type $AdS_4 \times H_2$ (and also a spherical one, for $AdS_4 \times S^2$). We find a 
fixed-point solution in the massive case, whereas for massless field configurations
we find a general solution. The latter has ``bad'' type
singularities: in order to resolve the singularities
we look for solutions with additional degrees of freedom in a similar
way to ref. \cite{Maldacena:2001yy}. This leads us to look for non-Abelian solutions.
Another possible way to resolve singularities is to find a black
hole solution where the singularity is covered by a horizon \cite{KLEBANOV}.
The aim of this paper is
to find the above mentioned gravity solutions, perform the up-lifting of
them to ten dimensions and, to study
their dual (twisted) gauge field theories. Massive type IIA supergravity is an
interesting theory which seems to be the most 
suitable arena for realizing de Sitter compactifications or Randall-Sundrum models \cite{LUIFA}
on manifolds with boundary. Therefore, the study of
the systems like the ones we deal with here could give some
ideas on general structures. 

Since in order to study the flow
between a higher dimensional theory and the resulting lower dimensional one it is necessary
to compactify some of the directions on a curved manifold, spinors have to be defined on 
non-flat spaces. Therefore, this requires to deal with a twisted gauge field theory living 
in such a manifold. Thus, in order to consider the dual field theories corresponding to certain 
supergravity solutions in curved
spaces, one has to study a twisted version of the corresponding supersymmetric theory. 
This new picture emerges from the impossibility in defining covariantly constant spinors 
in the aforementioned manifolds. Then, it is necessary to introduce a new term coming from 
the gauging of the normal bundle to the brane world-volume with the spin connection. 

We consider spacetimes with geometries given by $AdS_4 \times
\Sigma_2$ and $AdS_3 \times \Sigma_3$, such that 
their UV limit is an $AdS_6$-type spacetime. These lead to twisted field theories
since we assume $\Sigma_p$'s are curved manifolds. The decoupling limit
is taken in such a way that it preserves the 
volume of the compact manifold. Then, the twisted field theory
on the brane world-volume will be $R^{1, 4-p} \times \Sigma_p$, for D4-branes wrapping $\Sigma_2$ and
$\Sigma_3$. The field theory in this decoupling limit will not be sensitive to the
global geometry of $\Sigma_p$, remaining its effect
local and determining the twisted field theory
\cite{Maldacena:2000mw}. For energies much lower than the radius of
the compact manifold the theory will be a 3 (or 2) dimensional field theory, respectively.

This paper is organized as follows. 
In section 2 we review the basic formalism and set-up of the six-dimensional Romans' theories. 
Section 3 is devoted to the study of the gravity duals of 
superconformal field theories in 3 and 5 dimensions.
We focus on a non-Abelian solution, which will be up-lifted to massive
type IIA string theory. It can be interpreted as a D4-D8 system 
where the D4-brane is wrapped on 2- and 3-cycles. Section
4 begins with the up-lifting for the massless case to 
type IIA theory (even when expected this is another new result of
this work). Then,
we study the corresponding twisted gauge field theories. We discuss the
relation between our solutions and the previous ones \cite{Acharya:2000mu}, regarding the structure of their
actions in the string frame. Finally, we present our conclusions.

~


\section{The Romans' theories in 6 dimensions}


In this section we review the six-dimensional gauged  
${\cal {N}}=4$ supergravity constructed by Romans~\cite{Romans:1986tw},
whose conventions we follow. The theory consists of a graviton $e_\m^\a$, 
three $SU(2)$ gauge potentials $A_\m^I$, an Abelian potential  ${\cal A}_\m$, a
two-index tensor gauge field $B_{\m\n}$, a scalar $\phi$, four
gravitinos $\psi_{\m\, i}$ and four gauginos $\c_\m$. The bosonic
Lagrangian is 
\ba
e^{-1}\,{\cal L} &=& -\frac{1}{4} \, R +\frac{1}{2} \, 
(\partial^\mu\phi) \, (\partial_\mu\phi) - \frac{1}{4} \, 
{\rm e}^{-\sqrt{2}\f}\, ({\cal H}^{\mu\nu} \, {\cal H}_{\mu\nu} + 
F^{I \, \mu\nu} \, F^I_{\mu\nu})\nn\\
& &  + \frac{1}{12}\,{\rm e}^{2\sqrt{2}\f}\,G_{\m\n\r}\,G^{\m\n\r} + 
\fr18\,(g^2\,{\rm e}^{\sqrt{2}\f} + 4 g\,m\,{\rm e}^{-\sqrt{2}\f} -
m^2\,{\rm e}^{-3\sqrt{2}\f})\nn\\
& & - \frac{1}{8} \,e\,\varepsilon^{\m\n\r\s\t\k}\, B_{\m\n} \,
({\cal F}_{\r\s}\,{\cal F}_{\t\k} + m \, B_{\r\s} \, {\cal F}_{\t\k} +
\frac{1}{3} \, m^2 \, B_{\r\s} \, B_{\t\k} + F^I_{\r\s} \, F^I_{\t\k}),
\label{romansmassive}
\ea
where $e$ is the determinant of the vielbein, $g$ is the $SU(2)$ coupling constant, 
$m$ is the mass parameter associated with the two-index tensor field
$B_{\m\n}$ and
$\varepsilon_{\m\n\r\s\t\k}$ is a Levi-Civita tensor density. The
Abelian field strength ${\cal F}_{\m\n}$, the non-Abelian one $F_{\m\n}^I$, the three-form 
$G_{\m\n\r}$ and the field ${\cal H}_{\m\n}$ are given by
\ba
{\cal F}_{\m\n}&\equiv& \pa_\m{\cal A}_\n - \pa_\n{\cal A}_\m \,\,\, , \nn\\
F_{\m\n}^I &\equiv& \pa_\m A_\n^I - \pa_\n A_\m^I + g \,
\e^{IJK}\,A^J_\m\,A^K_\n \,\,\, , \nn\\
G_{\m\n\r} &=& 3\,\pa_{[\m}B_{\n\r]} \,\,\, , \nn\\
{\cal H}_{\m\n}&\equiv& {\cal F}_{\m\n} + m\,B_{\m\n} \,\,\, , 
\ea
respectively. The supersymmetry transformations for the gauginos and gravitinos are 
\ba
\d\c_i &=& \left( \fr1{\sqrt{2}}\, \g^\m \, 
\pa_\m\f + A \, \g_7 - \fr1{12} \,{\rm e}^{\sqrt{2}\f}\g_7\,\g^{\m\n\r}\, 
G_{\m\n\r}\right) \e_i + \fr1{2\sqrt{2}}\,\g^{\m\n}\,
({\hat{H}}_{\m\n})_i^{\,\,\, j} \, \e_j \,\,\, ,\label{gautransf}\\
\d\psi_{\m\, i} & = & \left( \nabla_\m  + T \, \g_\m \, \g_7
 - \frac{1}{24}\,{\rm
e}^{\sqrt{2}\f}\,\g_7\,\g^{\n\r\s}\,G_{\n\r\s}\,\g_\m 
\right) \, \e_i  \nonumber \\
&  & + \left( g \, A_\mu^I \, (T^I)_i^{\,\,\, j}
-\frac{1}{4\sqrt{2}} \, (\gamma_\mu^{\,\,\,\, \nu\rho} - 6 \, \delta_\mu^{\,\,\,\, \nu} \, \gamma^\rho)
({\hat{H}}_{\nu\rho})_i^{\,\,\, j} \right) \epsilon_j \,\,\, ,
\label{gratransf}
\ea
where $A$, $T$, and $\hat{H}$ are defined as follows
\ba
A &\equiv& \frac{1}{4\sqrt{2}} \, ( g \, \rme^{\frac{\phi}{\sqrt{2}}}
- 
3 \, m \, \rme^{\frac{-3\phi}{\sqrt{2}}}),\;\;\;\;\;\;\;\;\;\; 
T \equiv - \frac{1}{8\sqrt{2}} \, (g \,
e^{\frac{\phi}{\sqrt{2}}} + m \, \rme^{\frac{-3\phi}{\sqrt{2}}}) \,\,\,
,\label{atdef}\\
({\hat{H}}_{\mu\nu})_i^{\,\,\, j} &\equiv&
\rme^{-\frac{\phi}{\sqrt{2}}} \, \left( \fr12 {\cal H}_{\mu\nu} \, 
\delta_i^{\,\,\, j} + \gamma_7 \,
F_{\mu\nu}^I \, (T^I)_i^{\,\,\, j} \right) \,\,\,.
\label{hdef}
\ea
The gauge-covariant derivative ${\cal D}_\m$ acting on the Killing spinor is 
\beq
{\cal D}_\m\,\e_i = \nabla_\m\,\e_i + g\,A^I_\m\,(T^I)_i^{\;\;j}\,\e_j \,\,\, ,
\eeq
with
\beq
\nabla_\mu \e_i \equiv (\partial_\mu+\frac{1}{4} \, 
\omega^{\, \, \, \, \alpha \beta}_\mu  \,
      \gamma_{\alpha \beta} ) \, \e_i \,\,\, ,
\eeq
where $\omega^{\,\,\,\, \alpha \beta}_\mu$ is the
spin connection. Indices $\a, \b$ are tangent space (or flat) indices,
while $\m, \n$ are spacetime (or curved) indices. The 
$\gamma_{\alpha\beta\cdots}$ are the six-dimensional Dirac
matrices,
\[ \gamma_{\a_1 \cdots\a_n}=\frac{1}{n\,!} \,
\gamma_{[\a_1}\,\cdots\,\gamma_{\a_n]},\;\;\;\;\;\; n = 1, \cdots, 6.\]
The equations of motion of the Lagrangian~(\ref{romansmassive})
are
\ba
R_{\m\n} &=& 2\,\pa_\m\f\,\pa_\n\f + \fr18\,g_{\m\n}\,(g^2\,{\rm
e}^{\sqrt{2}\f} + 4\,g\,m\,{\rm e}^{-\sqrt{2}\f} - m^2\,{\rm
e}^{-3\sqrt{2}\f})\nn\\
& & + {\rm e}^{2\sqrt{2}\f}(G_\m^{\;\;\r\s}\,G_{\n\r\s} -
\fr16\,g_{\m\n}\, G^{\r\s\t}\,G_{\r\s\t})\nn\\
& & - 2\,{\rm e}^{-\sqrt{2}\f}({\cal H}_\m^{\;\;\r}\,{\cal H}_{\n\r} -
\fr18\,g_{\m\n}\,{\cal H}_{\r\s}\,{\cal H}^{\r\s})\nn\\ 
& & - 2\,{\rm e}^{-\sqrt{2}\f}(F_\m^{I\;\r}\,F^I_{\n\r} -
\fr18\,g_{\m\n}\,F_{\r\s}^I\,F^{I\;\r\s}),
\label{einsteineq}\\
\Box\f &=& \fr1{4\sqrt{2}}\,(g^2\,{\rm e}^{\sqrt{2}\f} - 4\,m\,g\,{\rm
e}^{-\sqrt{2}\f} + 3\,m^2\,{\rm e}^{-3\sqrt{2}\f}) + \fr1{3\sqrt{2}}\,{\rm
e}^{2\sqrt{2}\f}\, \nn\\ & & \times \,  G^{\m\n\r}\,G_{\m\n\r} 
+ \fr1{2\sqrt{2}}\,{\rm e}^{-\sqrt{2}\f}\,({\cal H}^{\m\n}\,{\cal
H}_{\m\n} + F^{I\;\m\n}\,F_{\m\n}^I),
\label{scalareq}\\
{\cal D}_\n\,({\rm e}^{-\sqrt{2}\f}\,{\cal H}^{\n\m}) &=&
\fr16\,e\,\varepsilon^{\m\n\r\s\t\k}\,{\cal H}_{\n\r}\,G_{\s\t\k},
\label{abelianeq}\\
{\cal D}_\n\,({\rm e}^{-\sqrt{2}\f}\,F^{I\;\n\m}) &=&
\fr16\,e\,\varepsilon^{\m\n\r\s\t\k}\,F^I_{\n\r}\,G_{\s\t\k},
\label{nonabelianeq}\\ 
{\cal D}_\r\,({\rm e}^{2\sqrt{2}\f}\,G^{\r\m\n}) &=&
- m\,{\rm e}^{-\sqrt{2}\f}\,{\cal H}^{\m\n} - 
\fr14\,e\,\varepsilon^{\m\n\r\s\t\k}\,({\cal H}_{\r\s}\,{\cal
H}_{\t\k} + F^I_{\r\s}\,F^I_{\t\k})\label{2formeq}.
\ea
Depending upon the values of the gauge coupling and mass parameter,
there are five distinct theories: 
${\cal N} = 4^+ \,\,\,$ (for $g > 0, m>0$),$\,\,\,$ ${\cal N} = 4^-\,\,\,$ (for $g <
0, m > 0$),$\,\,\,$  ${\cal N} = 4^g\,\,\,$ (for $g > 0, m = 0$),$\,\,\,$  
${\cal N} = 4^m\,\,\,$ (for $g = 0, m > 0$),$\,\,\,$ and ${\cal N} = 4^0\,\,\,$ 
(for $g = 0, m = 0$). The ${\cal N} = 4^g\,\,\,$ theory
coincides with a theory~\cite{Giani:1984dw} obtained by dimensional reduction
of gauged ${\cal N} = 2$ supergravity in seven dimensions, followed by truncation.
It was pointed out in~\cite{Romans:1986tw} that there is a dual version of
the ${\cal N} = 4^g$ theory, which has a similar field content but cannot be
obtained from the ${\cal N} = 4^g$ theory by a field redefinition. The $B_{\m\n}$ field
is replaced by a new tensor field $A_{\m\n}$, whose field strength is 
$\tilde{F}_{\m\n}^I$. In absence of the Abelian field strength, the bosonic Lagrangian of the dual
${\cal N} = \tilde{4}^g\,\,\,$ theory is
\be
\tilde{e}^{-1}\,\tilde{{\cal L}} = 
-\fr14\,\tilde{R} + \fr12\,\del_\m\tilf\del^\m\tilf - \fr14\,{\rm
e}^{-\sqrt{2}\tilf}\,\tilde{F}_{\m\n}^I\tilde{F}^{I\;\m\n} + 
\fr1{12}\,{\rm e}^{-2\sqrt{2}\tilf}\,
\tilde{F}_{\m\n\r}\tilde{F}^{\m\n\r} + 
\fr18\,\tilde{g}^2\,{\rm e}^{\sqrt{2}\tilf},
\label{n4tilde}
\ee
where
\be
\tilde{F}_{\m\n\r}\equiv 3\,\left(\del_{[\m}\,A_{\n\r]} - 
F^I_{[\m\n}\,A^I_{\r]} -
\fr13\,g\,\epsilon^{IJK}\,A^I_\m\,A^J_\n\,A^K_\r\right).  
\label{duallag}
\ee
The Lagrangian~(\ref{n4tilde}) can be obtained from the
Lagrangian~(\ref{romansmassive}) by formally writing 
$\tilde{F}_{\m\n\r}$ as follows
\be
\tilde{F}_{\m\n\r} = \fr16\,{\rm
e}^{2\sqrt{2}\,\tilf}\,e\,\varepsilon_{\m\n\r\s\t\k}\,G^{\s\t\k} \,\,\, .
\label{dualization}
\ee
Since the Lagrangian~(\ref{n4tilde}) differs from ${\cal N} = 4^g\,\,$ theory
only by a sign of dilaton coupling to three-form field, in the
absence of the three-form field ${\cal N} = 4^g\,\,$ and ${\cal N} =
\tilde{4}^g\,\,\,$ are identical. Furthermore, the
Eq.~(\ref{dualization}) tells us that any solution of ${\cal
N}=4^g\,\,$ theory with the two-form fields not being excited, which
is also a solution of ${\cal N} = \tilde{4}^g\,\,$ theory, can be
up-lifted into higher dimensional theory. We will consider the
ten-dimensional interpretation of ${\cal N} = 4^g\,\,$ theory in
section 4.


\section{Duals of 5 and 3-dimensional SCFTs}


In this section we study the gravity duals of superconformal field theories in five
dimensions with 8 supercharges \cite{Intriligator:1997pq}, which in the IR flow to superconformal
field theories in three dimensions with 4 supercharges, which is ${\cal {N}}=2$ in three  
dimensions \cite{Aharony:1997bx,KENTARO}.

The spinors of $SO(1,4)$ are pseudoreal with four components. These
theories in 5 dimensions are related by compactifications
to ${\cal {N}}=2$ four-dimensional theories. They have an $SU(2)_R$
symmetry of automorphism algebra which can be associated with the gauged symmetry in
the gravitational set-up of \cite{Romans:1986tw}. The vector multiplet in five
dimensions has one real
scalar component, a vector gauge field and a spinor, while the
hypermultiplet has four real scalars and a fermion. This theory has
a Coulomb branch when the real scalar has a {\it vev}, and a Higgs branch when
the scalar in the hypermultiplet is excited. Our gravitational
system can be understood as follows. Suppose that we start from type I theory on $R^9
\times S^1$, and we consider N D5-branes, the gauge theory on the branes
is ${\cal{N}}=1$ $Sp(N)$ gauge field theory with one hypermultiplet in the
antisymmetric representation and 16 hypermultiplets in the
fundamental. After T-dualizing on $S^1$ we arrive at a type I'
configuration on $S^1/Z_2$
with two O8 planes, N D4-branes and 16 D8-branes. The hypermultiplet in the
antisymmetric representation is massless while the mass of the fundamental hypermultiplets
is given by the relative position of the D8-branes with respect to the
D4-brane. The theory on the D4-brane is $Sp(N)$ with one vector multiplet (whose
scalar component describes the Coulomb branch $R^+$) and hypermultiplets
whose first components describe the Higgs branch. The global symmetries of
the theory are $SU(2)_R \times SU(2)\times SO(2N_f)\times U(1)$, being the
first $SU(2)$ the R-symmetry (the supercharges and the scalars in the
hypermultiplets are doublets under this group). The second one is associated with
the hypermultiplet in the antisymmetric representation and  the rest is
associated with the hypermultiplets in the fundamental and instantons. When the 
D4-brane is in the origin of the Coulomb branch we have a fixed point and
the global symmetry is enhanced to $ SU(2)\times E_{N_f +1}$. The
gravitational system is given by $N_f$ D8-branes situated on
a $O_8$ plane, $16 - N_f$ D8-branes in another fixed plane and, a D4-brane
which can move between them. The position of the D4-brane is
parameterized by the scalar in the vector multiplet. If $<\phi>$ is not
zero we have a theory with a $U(1)$ symmetry with $N_f$ ``electrons'', on
the fixed planes, the theory recovers its $SU(2)$ R-symmetry and it will have $N_f$ quarks.    

The gravity theory is a fibration of $AdS_6$ over $S^4$, with isometries
$SO(2,5)\times SU(2)\times SU(2)$.
The six-dimensional $SU(2)$
gauged supergravity has  an $AdS_6$  vacuum solution. It can be up-lifted
to massive IIA theory \cite{Cvetic:1999un} 
leading to the configuration described
above \cite{Nieder:2000kc}. In our case we will consider solutions of 
Romans' theory which, when uplifted, will give ten dimensional 
configurations with $N_f =0$. These theories were very well 
studied in \cite{Cachazo:2000ey} in the
context of algebraic geometry. There the authors studied the moduli spaces of 
type I' string and Heterotic string. 
It will be interesting to understand the case of
$N_f \neq 0$ and see whether there is a gravity solution dual to the
theory with enhanced symmetry $E_1'$. It will be also useful to
clarify the role of D0-branes in our gravity solutions. In principle they
should correspond to excitations of the six-dimensional Abelian gauge
field. We will not consider these fields here. We leave this for a future work.

In order to analyze the flow of the theory from five
dimensions to a superconformal theory in three dimensions with four
supercharges, we would like to consider a configuration where the
geometry ``loses'' two dimensions at low energies.  
Therefore, the idea is the following: we start with a six-dimensional gauged supergravity 
theory having an $AdS_6$ vacuum and 8 preserved supercharges. This
vacuum solution is dual to the five-dimensional SCFT mentioned above
and its ten-dimensional interpretation is given by a D4-D8 system. 
When one moves to the IR of the gauge theory (flowing in the radial
coordinate on the gravity dual) two of the dimensions of the theory become very small
and no low energy massless modes are excited on this two-space. Therefore, effectively
the gauge theory is three-dimensional. Besides, since the D4-brane 
is wrapped on a curved surface, we need to twist the theory. The
effect of this twisting is the breaking of some supersymmetries, as it can be seen
from the projections of the Killing vectors.

Under the group $SO(1,4) \times SO(3)_R$ the fields on the D4-brane transform
as $(\bf{1},\bf{1})$, $(\bf{3},\bf{1})$ and $(\bf{4},\bf{2})$. In the
IR the metric is taken to have $SO(1,2) \times SO(2)_D$ symmetry. 
The twisting involve ``mixing'' the $SO(2)$
group of the metric with an $SO(2)$ included in $SO(3)_R$. This mixing is
responsible for the breaking of some supersymmetries because the only
spinors that can be defined on the curved manifold are those that have 
scalar properties on the curved part. In fact, it breaks $1/2$ of the 
supersymmetries of the five-dimensional SCFT.

Below we construct a gravity dual to the five-dimensional SCFT theory by
finding a solution of Romans' theory with excited $F^I_{\m\n}$ 
fields. Then, we find a fixed-point solution for the massive field configuration.
We postpone the study of the massless case to the  
next section.

Let us consider the metric ans\"{a}tze of the form 
\beq
ds^2 = \rme^{2\,f} \, (dt^2 - dr^2 - dz^2 - dv^2 ) -
\frac{\rme^{2\,h}}{y^2} \, (dx^2 + dy^2) \,\,\, ,
\eeq
\beq
ds^2 = \rme^{2\,f} \, (dt^2 - dr^2 - dz^2 - dv^2 ) -
\rme^{2\,h} \, (d\theta^2 + \sin^2 \theta\, d\varphi^2) \,\,\, ,
\eeq
for $AdS_4 \times H_2$  and $AdS_4 \times S^2$, respectively.
The only non-vanishing non-Abelian gauge potential components are taken to be
\beq
A^{(3)}_x = \frac{a}{y} \hspace{.3in},\hspace{.3in}  A^{(3)}_\varphi = - a\cos\theta \,\,\, ,
\eeq
in the hyperbolic and spherical cases, respectively. 
One can treat both cases together by
introducing a parameter $\lambda$ that can take two values
\eqn{}{\lambda=+1 \to H_2 , \;\;\;\;\;\; \lambda=-1 \to S^2 \,\,\, .}
We impose the following projections
\eqn{projections}{\gamma_{45}(T^{(3)})_i^{\, j} \epsilon_j = \frac{\lambda}{2}
\epsilon_i, \;\;\;\;\;\; \gamma_2\gamma_7\epsilon_i= \epsilon_i  
\,\,\, .}
A solution must satisfy the equations obtained by setting to zero the supersymmetry transformations
for gauginos and gravitinos. After some algebra, they lead to
\beq
\varphi^\prime = \frac{\rme^f}{4\sqrt{2}}[g\, \rme^\varphi - 
3\,m\, \rme^{-3\varphi} + 2\, a\,\lambda \rme^{-2 h -\varphi}] \,\,\, ,
\label{martinn1}
\eeq
\beq
h^\prime = \frac{\rme^f}{4\sqrt{2}}[-g\, \rme^\varphi - m\, \rme^{-3\varphi} + 
6\,a\,\lambda\,\rme^{-2 h -\varphi}] \,\,\, ,
\label{martinn2}
\eeq
and 
\beq
f^\prime = -\frac{\rme^f}{4\sqrt{2}}[g\, 
\rme^\varphi + m\, \rme^{-3\varphi} + 2\, a\,
\lambda\,\rme^{-2 h -\varphi}] \,\,\, ,
\label{martinn3}
\eeq
where $\varphi \equiv \phi/\sqrt{2}$.

A fixed-point solution of these equations is given by
\ba
\rme^{4\varphi} & = & \frac{2 m}{g} \, , \nn \\
\rme^{-2h} & = & \sqrt{\frac{g\,m}{8}} \frac{1}{a\lambda} \, , \nn \\
\rme^{- f(r)} & = & \frac{g\,\rme^\varphi}{2\sqrt{2}}\,r \, .
\label{solfixed}
\ea  
In fact, it only happens if $\lambda \, a > 0$.
This solution satisfies the second order equations derived
from the starting six-dimensional Lagrangian.

We can analyze the structure of the solutions of the system above near the
UV of the theory ($r=0$). Indeed, we can see that an expansion leads to
the following behavior for the different fields
\eqn{}{f\approx - \log(r) + c_1 r^2 +..., \;\;\;\; g\approx - \log(r) + c_2
r^2 +...,\;\;\;\; \phi\approx  c_3 r^2 +... \
} 
(where $c_i$'s are constants). The interpretation of these expansions \cite{Balasubramanian:1999sn}
is that the scalar field $\phi$ 
describes the coupling of the five-dimensional field theory to an
operator of conformal weight $\Delta=3$ and mass squared $m^2=-6$. 
The operator is not the highest component of its supermultiplet,
thus it is a deformation that breaks some supersymmetry.

The ten-dimensional metric corresponding to the fixed-point solution
described above is easily obtained by using the results of \cite{Cvetic:1999un},
\eqn{}{X= \left(\frac{2m}{g}\right)^{1/4}\, ,\,\,\, \Delta = g\,X\,\cos^2\chi
+ 3\,m\,X^{-3}\,\sin^2\chi \,\,\, ,}
\ba
& & ds^2_{10}=(3\,m\,g^2)^{-1/8}\,(\sin\chi)^{\frac{1}{12}}
X^{1/8}\left[\frac{\Delta^{3/8}\,\rme^{2f_0}}{2\,r^2} (-dt^2 + dr^2 +
dv^2 + dz^2) \right.\nonumber\\
& &\left. + \frac{\sqrt{2}\Delta^{3/8}}{\sqrt{g^3\,m}\,y^2}(dx^2 + dy^2 )
+ \Delta^{3/8}\,\sqrt{2\,m\,g^3}\,d\chi^2\right.\nonumber\\
& &\left. +\frac{\Delta^{-5/8}}{g\,X}\,\cos^2\chi \left(\left(\sigma_1 -\frac{dx}{y}\right)^2
+\sigma_2^2 + \sigma_3 ^2\right)\right] \,\,\, .
\ea
The metric has symmetries of the form $SO(2) \times SO(3)$.
The corresponding expression for the RR fields and the ten-dimensional
dilaton can be read from \cite{Cvetic:1999un} and we do not quote it here.

Although we could not find an exact solution, 
we can carry out a similar analysis to \cite{Acharya:2000mu}. Actually, 
the two cases $\lambda=\pm 1$ can be gathered again. Thus defining 
$F\equiv u^2 e^{-2 \varphi}\;,\;u\equiv e^{2 h(r)}$ 
and using Eqs.(\ref{martinn1}), (\ref{martinn2}) we obtain
the following differential equation
\beq
\frac{d F}{d u} = \left(\frac{ 3 g u^4 - m F^2 - 10
a \lambda u F}{ g u^4 + m F^2- 6 a \lambda u F} \right) \frac{F}{u} \,\,\, .
\label{ORBITS}
\eeq
For the case $\lambda=+1$ (hyperbolic plane), we can solve this equation in 
the approximations when $u\to \infty$ and $u\to 0$. In fact, in the UV one has 
\eqn{}{F\approx  u^2 + ..., \,\,\,\,\;\; \rme^{-2\varphi}\approx 1+... ,\;\;   }
and the metric results to be the $AdS_6$ limit of our original metric.
In the IR we have 
\eqn{}{F\approx F_0 u^{5/3} + ..., \;\;\,\,\,\, e^{-2\varphi}\approx F_0 u^{-1/3}+... 
}
and the metric is 
\eqn{}{ds_6^2= \frac{1}{u^{1/3}}(dt^2 - dz^2 - dv^2) - \frac{u}{y^2}(dx^2 + dy^2)
- \frac{2}{9 \, F_0 \, a \, u^{1/3}}du^2 \,\,\, .   }

\vspace{0.5cm}
\begin{center}
\epsfig{file=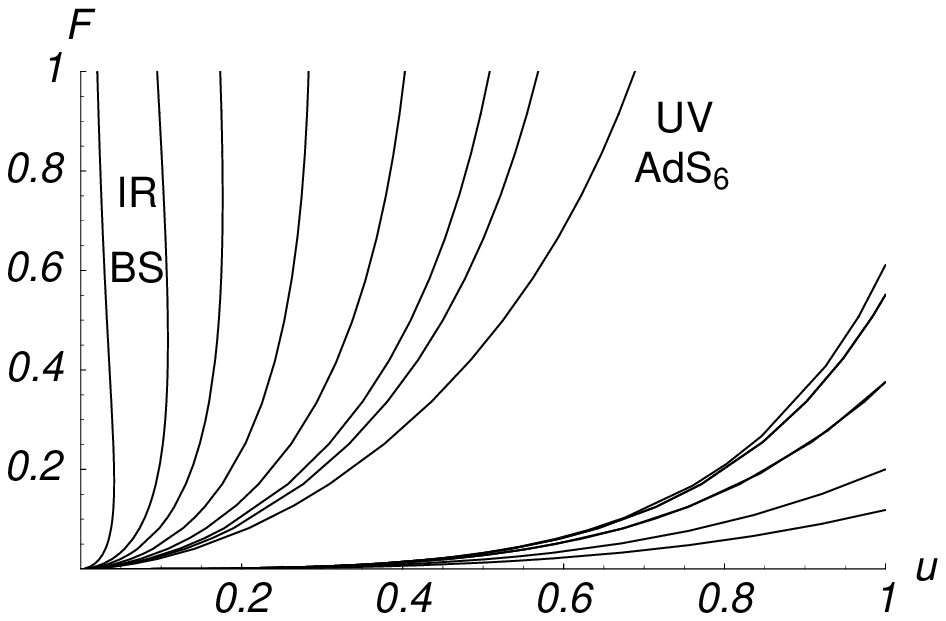, width=9cm}
\end{center}
\vspace{0.5cm}
\baselineskip=13pt
{\small{Figure 1: Orbits for the case $AdS_4 \times H_2$. In the UV limit $F$ behaves as $u^2$ leading to 
an $AdS_6$-type region. On the other hand, it flows to ``bad'' singularities (BS) in the IR.
We have used $g = 3 m$ with $m=\sqrt{2}$, and $a=1$.}}

\vspace{0.5cm}

\baselineskip=15.5pt

When up-lifted the singularities
turn out to be of the ``bad''type. They correspond to the
curves flowing to the origin of the plots in figure 1. On the other hand, 
in both cases of the hyperbolic plane and sphere there are other kinds of ``bad'' 
singularities that correspond to the orbits going like $F \approx \frac{1}{u}$ as 
$u$ approaches to zero. Particularly, one can see this kind of behavior in
figure 2. For the two-sphere we have the set of orbits depicted in figure 2.

This also constitutes a part of the IR behavior of the flow. As we can see in figures
1 and 2, there is a smooth solution interpolating between the UV and the IR limits 
of the system, thus
realizing our initial set up. The fixed-point solution is out of the range of this plot. 

\newpage

\vspace{0.5cm}
\begin{center}
\epsfig{file=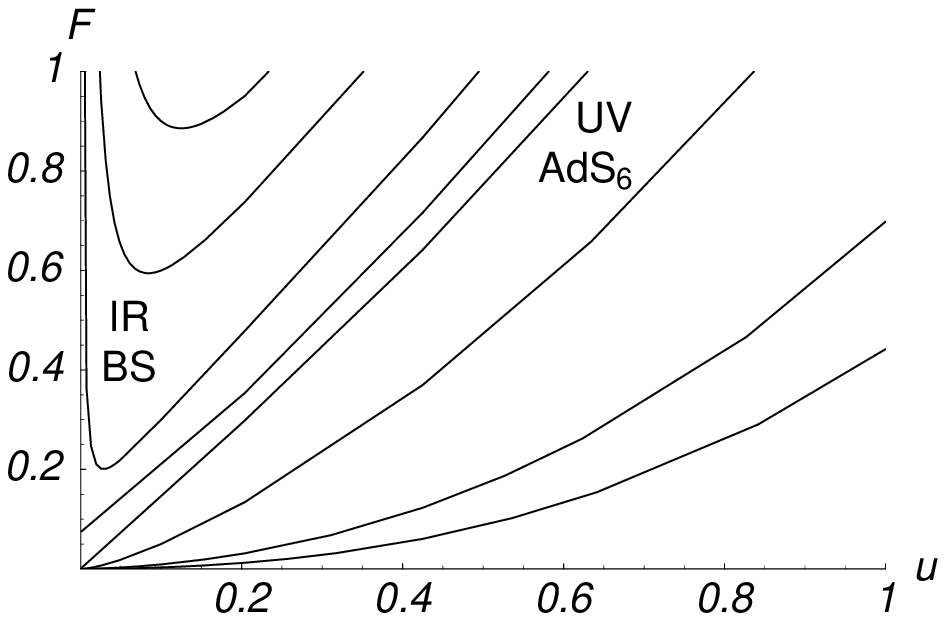, width=9cm}
\end{center}
\vspace{0.5cm}
\baselineskip=13pt
{\small{Figure 2: We show the behavior of the orbits for the case $S^2$.
Similar comments as in the previous case are also valid here. For small $u$-values
we can see the region corresponding to $F \approx \frac{1}{u}$.}}

\vspace{0.5cm}

\baselineskip=15.5pt

In summary, we have found a fixed-point solution for the hyperbolic plane. 
In the UV both geometries flow to their corresponding asymptotic
$AdS_6$. In the IR limit all the singularities in both geometries are of 
``bad'' type according the criterion of ref. \cite{Maldacena:2000mw}.
Similar behaviour is expected for the sphere.
Note that the solutions studied here only represent a special case of
solutions that, in general, have the form of a Laurent series. The fact
that these solutions 
have non-acceptable singularities can be easily seen from the
construction of an effective potential corresponding to this effective four
dimensional system, and following the analysis outlined by Gubser in
ref. \cite{Gubser:2000nd} one can see that these kinds of solutions
will not be interpretable as the IR of a gauge theory.
However, this does not exclude the possibility of finding a different
solution that could have an interpretation as the transcendental solution
found in ref. \cite{Acharya:2000mu}.

\subsection{A non-Abelian solution in massive type IIA theory}

In this section we describe a non-Abelian solution in the
massive IIA theory. Therefore, we consider a gravity dual of a 
five-dimensional SCFT with 8 supercharges that flows through the dimensions
to a two-dimensional (1,1) CFT with two supercharges at the IR. Its
field content is given by a real scalar, a gauge field and the
corresponding fermionic partner. The idea is to find a compactification
from six-dimensional supergravity on $AdS_3 \times H_3$. Under
$SO(1,4)\times SO(3)_R$ the charges transform as $(\bf{4},\bf{2})$ and
$(\bf{\bar{4}},\bf{2})$. After the twisting $ SO(1,4)\times SO(3)_R \to
SO(1,1)\times SO(3) \times SO(3)_R \to SO(1,1) \times SO(3)_D$ we are left
with two supercharges which are scalars under the diagonal subgroup. $SO(3)_D$
results from the identification  of the normal bundle of the
D4-brane world-volume with the spin bundle of $\Sigma_3$.
They are the two ``twisted'' supercharges that survive the twisting process. The field
content of the theory can be obtained along the same lines and the result is
the one mentioned above.

We believe that this solution is
the first example of a non-Abelian solution of massive type IIA theory.
It is similar to the solution obtained in
\cite{Acharya:2000mu} for massive seven-dimensional gauged
supergravity. In that case the solution up-lifts to M theory. Therefore,
this  suggests some connection between theories in the line of ref. 
\cite{Lavrinenko:1999xi} that merits further exploration.
 
Our configuration is
\eqn{}{ds^2 = \rme^{2f}(dt^2 - dr^2 - du^2 )- \frac{\rme^{2 h}}{y^2}(dx^2 + dz^2 
+ dy^2),\;\; A_x^{(1)}= \frac{a}{y},~ A_z^{(3)}= \frac{b}{y} \, ,
}
we can also consider an $AdS_3 \times S^3$, for $\lambda = -1$. 
The projections (all the indexes are plane indexes) are given by 
\eqn{}{\gamma_{65} (T^{(3)})^j_i\epsilon_j = \frac{1}{2}\epsilon_i,\;\;
\gamma_{64} (T^{(2)})^j_i\epsilon_j = \frac{1}{2}\epsilon_i,\;\;
\gamma_{45} (T^{(1)})^j_i\epsilon_j = \frac{1}{2}\epsilon_i,\;\;
\gamma_7\gamma_2 \epsilon_i = \epsilon_i \,\,\, ,
}
we use $\xi= e^{\frac{\phi}{\sqrt{2}}}$.
The BPS equations are
\eqn{}{a\,b = \frac{1}{g} \,\,\, ,
}
\eqn{}{h^\prime= - \frac{\rme^{f}}{4 \sqrt{2}} [ g \xi + m \xi^{-3} - 10 \lambda \,a
\xi^{-1}\,\rme^{-2 h}] \,\,\, ,
}
\eqn{}{\phi^\prime = \frac{\rme^f}{4 \sqrt{2}} [ g \xi  - 3 m \xi^{-3} +
6 \lambda \, a\, \xi^{-1}\, \rme^{-2 h}] \,\,\, ,
}
\eqn{}{f^\prime = - \frac{\rme^f}{4 \sqrt{2}} [g \xi + m \xi^{-3} + 6 \lambda \,a\,
\xi^{-1}\,\rme^{-2 h}] \,\,\, .
}
A fixed-point solution is obtained and it reads as it follows
\eqn{}{\rme^{-2 h}= \frac{\sqrt{g^3\,m}}{2\sqrt{6}}, \,\,\,\;\; \xi =
\left(\frac{3m}{2 g}\right)^{1/4} \,\,\, .}

This configuration can be uplifted to ten dimensions
(massive IIA theory) using the results in \cite{Cvetic:1999un}
\eqn{}{X= e^{\phi/\sqrt{2}}, \Delta = g\,X\,\cos^2\chi + 3\,m\,X^{-3}\,\sin^2\chi \,\,\, ,}
\ba
& & ds^2_{10}=\left(\frac{1}{3\,m\,g^2}\right)^{1/8}\,(\sin\chi)^{\frac{1}{12}}
X^{1/8}\left[\frac{\Delta^{3/8}}{2\,r^2} (-dt^2 + dr^2 +
dv^2)\right.\nonumber\\
& & \left. + \frac{\rme^{2h}}{2\,y^2}\,\Delta^{3/8}\,(dx^2 +dz^2+
dy^2) + g^2\,\xi\,\Delta^{3/8}\,d\chi^2 \right.\nonumber\\
& & \left. + 
\frac{\Delta^{-5/8}}{g\,X}\,\cos^2\chi \left( \left(\sigma_1 - \frac{dx}{y}\right)^2 
+\sigma_2^2 + \left(\sigma_3 -\frac{dz}{y}\right)^2\right)\right] \,\,\, ,
\ea
where $\sigma_i$ are the three left-invariant forms in the three-sphere. 
We can see that the metric has an $SO(2) \times SO(3)$ invariance.

Since we are not able to integrate the above BPS system, we will
repeat the analysis of the singularities as we did in the previous section.
Again, both cases (hyperbolic plane and sphere) are treated together by using
the parameter $\lambda = \pm 1$. Proceeding as above we get the following
first order differential equation
\eqn{}{
\frac{d F}{d u} = \left( \frac{3 g u^4 - m F^2 -14 a \lambda u F}{g u^4 + m F^2 -
10 a \lambda u F} \right) \frac{F}{u}\,\,\, . 
}
When $\lambda = +1$ we solve this equation in two approximations,
$u \to \infty$ and $u \to 0$. The large-$u$ approximation leads to the
corresponding asymptotic $AdS_6$ whereas the other case gives
  
\eqn{}{F\approx  F_0 \frac{1}{u} + ..., \,\,\,\,\;\; \rme^{-2\varphi}\approx 
\frac{F_0}{u^3}+... ,\;\;   
}
\eqn{}{F\approx - 2 \frac{a}{m}u + ..., \;\;\,\,\,\, e^{-2\varphi}\approx 
-\frac{2 a}{m u}+... 
}
\eqn{}{F\approx F_0 u^{7/5} + ..., \;\;\,\,\,\, e^{-2\varphi}\approx 
F_0 u^{3/5}+... \,\,\, .
}
Note that all the singularities are of the ``bad'' type in the IR. 

\vspace{0.5cm}
\begin{center}
\epsfig{file=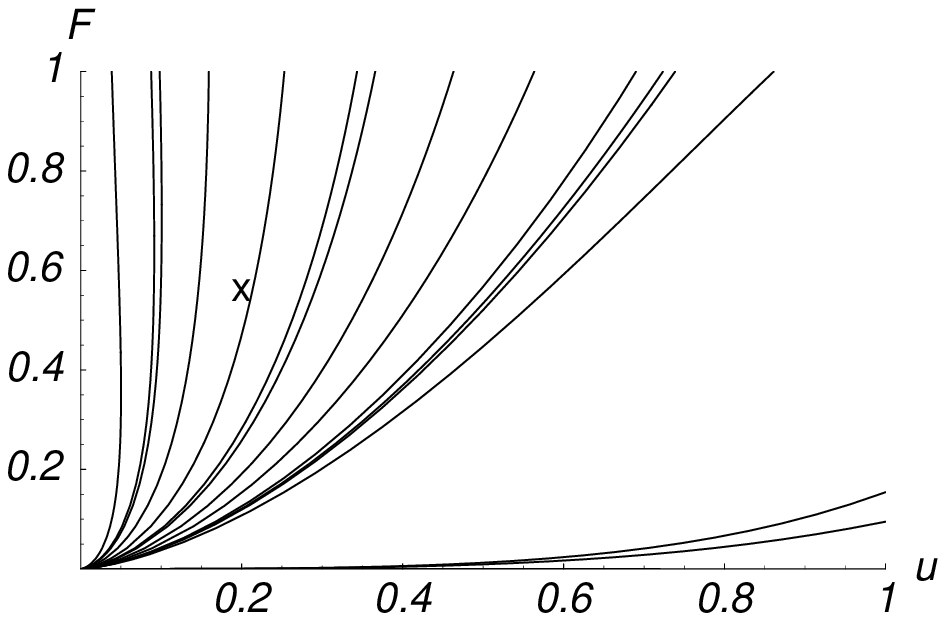, width=9cm}
\end{center}
\vspace{0.5cm}
\baselineskip=13pt
{\small{Figure 3: Behavior of the orbits for the non-Abelian massive case.
The $AdS_6$-type region is at the UV, {\it i.e.}, for large
$u$ and $F$-values. In the IR it flows to ``bad'' singularities (BS). The
cross indicates the fixed point-solution aforementioned. We again set $g = 3 m$,
$m= \sqrt{2}$, and $a=b=\lambda=1$.}}

\vspace{0.5cm}

\baselineskip=15.5pt

Here we should make a similar comment as the one in the previous section:
Gubser's criteria tells us to construct the effective potential of the
three-dimensional gravity theory. With this potential we can see that the solutions in the
form of a Laurent series showed above will not have an interpretation as
the IR limit  of a gauge theory. 

\section{The massless case}

Consider six-dimensional gauged supergravity with the mass
parameter taken to be zero ($m=0$). We would like to
find a ten-dimensional Kaluza-Klein ansatz which allows us to perform the up-lifting
of certain solutions of the six-dimensional Romans' theory. In this section, we will show that 
the ${\cal{N}} = \tilde{4}^g$ theory can be embedded in type IIA theory.

Thus, we start with type IIA theory in ten dimensions and set the two- and three-form field
strengths to zero.  After the truncation the bosonic Lagrangian is
\begin{equation}
\hat{{\cal L}}_{10} = \hat{R}*1 - \frac{1}{2} *d\hat{\phi}\wedge d\hat{\phi} -
\frac{1}{2}{\rm e}^{-\frac{1}{2}\hat{\phi}}*\hat{F}_{(4)}\wedge \hat{F}_{(4)} \,\,\, .
\end{equation}
The ten-dimensional equations of motion are
\begin{eqnarray}
d\hat{*}d\hat{\phi} &=& - \frac{1}{4}{\rm
e}^{-\frac{1}{2}\hat{\phi}}\,\hat{*}\,\hat{F}_{(4)}\wedge
\hat{F}_{(4)},\nonumber\\
d({\rm e}^{-\frac{1}{2}\hat{\phi}}\hat{*}\hat{F}_{(4)}) &=&
0,\nonumber\\
\hat{R}_{\mu\nu} &=& \frac{1}{2}\,\partial_\mu\hat{\phi}\partial_\nu\hat{\phi} + 
\frac{1}{12}{\rm
e}^{-\frac{1}{2}\hat{\phi}}\,(\hat{F}_{\mu\r\s\t}\,
\hat{F}_{\nu}^{\;\;\r\s\t} -
\frac{3}{32}g_{\mu\nu}\,\hat{F}_{(4)}^2),\nonumber\\
\hat{F}_{(4)}\wedge \hat{F}_{(4)} &=& 0 \,\,\, .
\end{eqnarray}
Notice that the last equation is
a constraint resulting from the fact that we set $\hat{F}_{(3)} = \hat{\cal
F}_{(2)} = 0$.
 
Reducing the theory on a circle, we get a nine-dimensional
theory whose bosonic sector consists of a graviton, two scalars, a
three-form field strength and a two-form field strength. One can
consistently truncate this theory to a theory whose bosonic content
includes a graviton, a scalar and a three-form field strength. The
ans\"atze for the metric, scalar and three-form fields are
\ba
d\hat{s}_{10}^2 &=& {\rm e}^{-\frac{5}{8\sqrt{14}}\bar{\f}} d\bar{s}_9^2 +
{\rm e}^{\frac{35}{\sqrt{14}}\bar{\f}}\,dZ^2,\nn\\
\hat{\phi} &=& \frac{\sqrt{14}}{8}\,\bar{\f},\nn\\
\hat{F}_{(4)} &=& \bar{F}_{(3)}\wedge dZ \,\,\, .
\ea
The resulting nine-dimensional Lagrangian and its equations of motion are
\begin{eqnarray}
\bar{{\cal L}} &=& \bar{R}*1 - \frac{1}{2}*d\bar{\phi}\wedge
d\bar{\phi} - 
\frac{1}{2}{\rm e}^{-\frac{4}{\sqrt{14}}\bar{\phi}}*\bar{F}_{(3)}\wedge
\bar{F}_{(3)},\nonumber\\
0 &=& d({\rm e}^{-\frac{4}{\sqrt{14}}\bar{\phi}}
\bar{*}\bar{F}_{(3)}),
\nonumber\\
\Box\bar{\phi} &=& -\frac{1}{3\sqrt{14}}{\rm
e}^{-\frac{4}{\sqrt{14}}\bar{\phi}}\bar{F}_{(3)}^2, \nn\\ 
\bar{R}_{\mu\nu} &=& \frac{1}{2}\partial_\mu\bar{\phi}\,\partial_\nu\bar{\phi}
+ \frac{1}{4}{\rm e}^{-\frac{4}{\sqrt{14}}\bar{\phi}} 
\left(\bar{F}_{\mu\r\s}\bar{F}_\nu^{\;\;\r\s} - 
\frac{2}{21}g_{\mu\nu}\bar{F}_{(3)}^2\right) \,\,\, .
\end{eqnarray}
Now, we compactify the 9-dimensional theory on
$S^3$ using the formulae in~\cite{Cvetic:2000dm}. The metric ansatz is
\begin{equation}
d\bar{s}_9^2 = X^{3/14} ds_6^2 + \frac{1}{4\,\tilg^2} X^{-2/7} 
\sum_{i=1}^3\, (\sigma^i - \tilg\,\tilde{A}^i_{(1)})^2,
\end{equation}
where $\sigma^i, i=1,2,3$ are left-invariant 1-forms of $SU(2)$
\ba
\s^1 &=& \cos\psi\,d\theta + \sin\psi\,\sin\theta\,d\varf,\nn\\
\s^2 &=&  -\sin\psi\,d\theta + \cos\psi\,\sin\theta\,d\varf,\nn\\
\s^3 &=& d\psi + \cos\theta\,d\varf,
\label{leftinva}
\ea
and $X = \exp(\sqrt{2}\,\tilde{\phi})$.
The ans\"atze for three-form field, its Hodge dual and the scalar field are
\begin{eqnarray}
{\rm e}^{\sqrt{\frac{7}{2}}\bar{\phi}} &=& X \,\,\, , \nonumber\\
\bar{F}_{(3)} &=& \tilF_{(3)} - 
\frac{1}{4\,\tilg^2}\,\tilh^1\wedge \tilh^2\wedge \tilh^3 
+ \frac{1}{4}\sum_{i=1}^3\,\tilF^i_{(2)}\wedge \tilh^i,\nonumber\\
{\rm e}^{-\frac{4}{\sqrt{14}}\bar{\f}}\,\bar{*}\,\bar{F}_{(3)} &=&
-\frac{1}{8\tilg^3}X^{-1}\,*\tilF_{(3)}\wedge \tilh^1\wedge
\tilh^2\wedge \tilh^3 + 2\,\tilg\,X^{1/2}\,\epsilon_{(6)}\nonumber\\
& & + \frac{1}{16\tilg^2}\,\epsilon_{ijk}\,X^{-1/2}\,*\tilF^{i}_{(2)}\wedge
\tilh^j \wedge \tilh^k,
\end{eqnarray}
where $\tilh^i \equiv \s^i - \tilg\,\tilA^i$. Substituting these ans\"atze into 
the nine-dimensional equations of motion we get
\begin{eqnarray}
d\tilF_{(3)} &=& \frac{1}{2}\,\sum_{i=1}^3\,\tilF_{(2)}^i\wedge
\tilF_{(2)}^i,\nonumber\\
d(X^{-1}\,*\tilF_{(3)}) &=& 0,\\
\label{6dimeq}
{\cal D}(X^{-1/2}\,*\tilF_{(2)}^i) &=& X^{-1}\,*\tilF_{(3)}\wedge
\tilF_{(2)}^i,\nonumber\\
d(X^{-1}\,*dX) &=& X^{-1}\,*\tilF_{(3)}\wedge \tilF_{(3)} +
\frac{1}{4}\,X^{-1/2}\,\sum_{i=1}^3\, *\tilF^i_{(2)}\wedge \tilF^i_{(2)} -
4\tilg^2\,X^{1/2}\,\epsilon_{(6)}.\nn
\end{eqnarray}
The ansatz and the Lagrangian that produce Eq.~(\ref{6dimeq}) are
\begin{eqnarray}
d\hat{s}_{10}^2 &=& {\rm e}^{\frac{\sqrt{2}}{8}\tilde{\phi}} ds_6^2 + 
\frac{1}{4\tilg^2}\,{\rm e}^{-\frac{3\sqrt{2}}{8}\tilde{\phi}}
\sum_{i=1}^3\,(\sigma^i - \tilg\,\tilA^i_{(1)})^2 + 
{\rm e}^{\frac{5\sqrt{2}}{8}\tilde{\phi}}dZ^2,\nonumber\\
\hat{F}_{(4)} &=& (\tilF_{(3)} -
\frac{1}{4\tilg^2}\,\tilh^1\wedge \tilh^2\wedge \tilh^3 + 
\frac{1}{4\tilg}\,\tilF^{i}_{(2)}\wedge \tilh^i)\wedge dZ,\nonumber\\
\hat{\phi} &=& \frac{1}{2\sqrt{2}}\,\tilde{\phi},\nonumber\\
\frac{1}{\sqrt{g}}{\cal L}_6 &=& \tilde{R} - 
\frac{1}{2}(\partial\tilde{\phi})^2 - 
\frac{1}{4}{\rm e}^{-\frac{1}{\sqrt{2}}\tilde{\phi}}\,
\tilF^{i}_{\mu\nu}\tilF^{i\;\mu\nu} +
4\,\tilg^2 {\rm e}^{\frac{1}{\sqrt{2}}\tilde{\phi}}
- \frac{1}{12}{\rm e}^{-\sqrt{2}\tilde{\phi}}\tilF_{(3)}^2 \,\,\, ,
\label{finalansatz}
\end{eqnarray}
where $\tilF_{(3)} = d \tilA_{(2)} + \frac{1}{4}(
\tilF^{i}_{(2)}\wedge \tilA^i_{(1)} - \frac{1}{6}
\tilg\,\epsilon_{ijk}\,\tilA_{(1)}^{i}
\wedge \tilA_{(1)}^{j}\wedge \tilA_{(1)}^{k})$. For an easy comparison
with the Romans' theory we make the following rescalings
\be
\tilde{g}_{\m\n}\rightarrow -\fr12\,g_{\m\n},\; \tilg\rightarrow
\fr12\,g,\; \tilA^i\rightarrow \sqrt{2}\,A^i,\;\tilf \rightarrow 
2\,\f,\;\tilF_3\rightarrow F_3 \,\,\, .
\ee
The Lagrangian and the ansatz in Eq.(\ref{finalansatz}) now become
\begin{eqnarray}
d\hat{s}_{10}^2 &=& -\frac{1}{2}{\rm e}^{\frac{\sqrt{2}}{4}\phi}\, ds_6^2 + 
\frac{1}{g^2}\,{\rm e}^{-\frac{3\sqrt{2}}{4}\phi}\,
\sum_{i=1}^3\,\left(\sigma^i - \frac{1}{\sqrt{2}}\,g\,A^i_{(1)}\right)^2 + 
{\rm e}^{\frac{5\sqrt{2}}{4}\phi}\,dZ^2,\nonumber\\
\hat{F}_{(4)} &=& \left(\,F_{(3)} -
\frac{1}{g^2}h^1\wedge h^2\wedge h^3 + 
\frac{1}{\sqrt{2}\,g}\,F^{i}_{(2)}\wedge h^i\,\right)\wedge dZ,\nonumber\\
\hat{\phi\baselineskip=20pt plus 1pt minus 1pt
} &=& \frac{1}{\sqrt{2}}\,\phi,\nonumber\\
e^{-1}\,{\cal L}_6 &=& -\fr14\,R + 
\frac{1}{2}(\partial\phi)^2 - 
\frac{1}{4}{\rm e}^{-\sqrt{2}\phi}\,
F^{i}_{\mu\nu}\,F^{i\;\mu\nu} +
\fr18\,g^2 {\rm e}^{\sqrt{2}\phi}
+ \frac{1}{12}{\rm e}^{-2\sqrt{2}\phi}\,F_3^2, 
\label{finalansatz1}
\end{eqnarray}
where $h^i = \s^i - \frac{1}{\sqrt{2}}\,g\,A^i$. The Lagrangian in
Eq.~(\ref{finalansatz1}) is precisely the ${\cal{N}}=\tilde{4}^g$ Romans' theory.

We can study a similar problem to the one in the previous section.
One can write down a similar ansatz for the metric, {\it i.e.} a
wrapped product of $AdS_4 \times \Sigma_2$, where $\Sigma_2$ is a an
hyperbolic plane or a sphere, as before.
The system of Eqs.(\ref{martinn1})-(\ref{martinn3}) 
can be studied in the massless case, {\it i.e.}
$m=0$, while $g$ and $a$ are non-vanishing quantities. We have the solution
\eqn{}{
f =- \varphi,  \,\,\,\,\,\,\, \varphi = 
\frac{g}{4\sqrt{2}}\,r + \frac{1}{8}\log(r) ,\;\;\,\,\,\,\, 
h = -\frac{g}{4\sqrt{2}}r + \frac{3}{8}\,\log(r) \,\,\, .
}

Therefore, the six-dimensional metric is given by
\eqn{badmetric}{ds^2 = \frac{\rme^{-\frac{g r}{2\sqrt{2}}}}{r^{1/4}}
[dt^2 - dr^2 - dz^2 - dv^2 -  r d\Omega^2_{\lambda}].
}

The metric has a singularity. One way to resolve it is to look for a
solution that has more degrees of
freedom excited such as
a non-Abelian solution. Another way to have an
acceptable gravity solution is to construct a black hole solution
such that the singularity is hidden by the horizon \cite{KLEBANOV}. 
In the next section
we will discuss a resolution of the singularity in the $S^2$ case, {\it i.e.},
when the parameter $\lambda =-1$ in Eqs.(\ref{martinn1}) to
(\ref{martinn3}).

~

\subsection{A non-Abelian solution}

We would like to present a non-Abelian solution
that resolves the ``bad'' singularity of the solution Eq.(\ref{badmetric}). 
This solution will be useful for the study of the gravity dual for the
three-dimensional ${\cal{N}}=2$ super Yang-Mills theory. Actually
this solution, being related to the solution in reference \cite{Maldacena:2001yy}
represents a smeared NS-fivebrane on $S^2$ after a T-duality. In the IR
the theory living on the brane will be ${\cal{N}}=2$ super Yang-Mills theory in three dimensions
plus Kaluza-Klein  modes that do not decouple.
The solution in the string frame is
\ba
ds^2 &=& 2\, [dt^2 - dr^2 - d\vec{x}_2^2 - \rme^{2\,h}(d\theta^2 +
\sin\theta^2 d\varf^2) ],\nn\\
F &=& -w^\prime \s^2 dr\wedge d\theta + 
w^\prime \sin\theta \s^1 dr\wedge d\varphi + (w^2 -
1)\,\s^3\,\sin\theta d\theta \wedge d\varphi,\nn\\
\phi_S &=& \frac{1}{2} \log\left(\frac{\sinh(r)}{R(r)}\right), 
\;\;\; w(r)=\frac{r}{\sinh(r)}, w^\prime = \frac{dw}{dr},\nn\\
\rme^{2\,h} &=& R(r)^2, R(r)=\sqrt{2r \coth(r) -
r^2/\sinh(r)^2 - 1} \,\,\, .
\ea
This is the same solution reported by Chamseddine and Volkov in  
\cite{Chamseddine:1997nm,Chamseddine:1998mc}.

The relations between the scalars and metric in Einstein frame and
string frame are
\[\varphi = -\fr12 \phi_S;\;\;\;\;\; g_{\mu\nu}(E) = 
{\rm e}^{\phi_S}g_{\mu\nu}(S).\] 
In Einstein frame the solution is
\begin{equation}
ds^2_E = 2 {\rm e}^{\phi_S}(-dt^2 + dr^2 + d\vec{x}_2^2) + {\rm
e}^{2\,G(r)}(d\theta^2 + \sin^2\theta d\varphi^2),
\end{equation}
where the dilaton is given by
\begin{eqnarray}
\phi_S &=&
\frac{1}{2}\log\left(\frac{\sinh(r)}{R(r)}\right),\nonumber\\
R(r) &=& \sqrt{2 \, r \, \coth(r) - w(r)^2 -1},\;\;\; {\rm e}^{2\,G} =
R(r)^2,\;\;\; w(r) = \frac{r}{\sinh(r)} \,\,\, .
\end{eqnarray}
The field strength components are
\begin{eqnarray}
F^1_{(2)} &=& w^{\prime} \sin\theta dr\wedge d\varphi,\nonumber\\
F^2_{(2)} &=& -w^{\prime} dr\wedge d\theta,\nonumber\\
F^3_{(2)} &=& (w^2-1)\sin\theta d\theta\wedge d\varphi \,\,\, .
\end{eqnarray}
Collecting all we have the following ten-dimensional solution:
\begin{eqnarray}
d\hat{s}_{10}^2 &=& {\rm
e}^{-\frac{3\sqrt{2}}{8}\phi}\left[-dt^2 + dr^2 + d\vec{x}_2^2 + {\rm
e}^{2G}(d\theta^2 + \sin^2\theta d\varphi^2) + \frac{1}{g^2}\,
\sum_{i=1}^3\,(\sigma^i - \frac{g}{\sqrt{2}}\,A^i_{(1)})^2\right]\nonumber\\ 
& & + {\rm e}^{\frac{5\sqrt{2}}{8}\phi} dZ^2\nonumber\\
\hat{F}_{(4)} &=& (-\frac{1}{g^2}h^1\wedge h^2\wedge h^3 +
\frac{1}{\sqrt{2}\,g}\,F^i_{(2)}\wedge h^i)\wedge dZ,\nonumber\\
\hat{\phi} &=& \frac{1}{\sqrt{2}}\phi = -
\frac{1}{4}\log\left(\frac{\sinh(r)}{R(r)}\right)\,\,.
\end{eqnarray}

~

~

\noindent
{\em A black hole solution in the massless theory}


~

The solutions of the massless six-dimensional Romans' theory, 
the string frame metric and 2-form ansatz are
\ba
ds^2 &=& \frac{f(r)}{r}dt^2 - \frac{1}{r\,f(r)}\,dr^2 - R^2\,(d\theta_1^2
+
\sin^2\theta_1\,d\varphi_1^2) - dx^2 - dy^2,\nn\\
F_2 &=& h(r)\,dr\wedge dt + \frac{R\,Q_m}{\sqrt{2}}\sin\theta_1\,
d\theta_1\wedge d\varphi_1,
\ea
where $f$ and $h$ are functions of $r$ while $\g$ and $Q_m$ are
constants.

The solution in the string frame is
\be 
h(r) = \frac{Q_e}{r^2},\;\;
f(r) = - M + \frac{2\,Q_e^2}{r} + \left(\frac{g^2}{2} + 
\frac{Q_m^2}{R^2}\right)r,\;\;\f(r) = \frac{1}{2}\,\log(r).
\ee
Similar solutions were previously discussed
 in~\cite{Cvetic:1999pu}. Our solution has a different geometry
and can carry both electric and magnetic charges.

One can up-lift the above solution to type IIA theory, and the results are
\ba
d\hat{s}_{10}^2 &=& -\frac{1}{2}\,r^{-1/8}\, \times \nn\\
& & \left(
\frac{f(r)}{\sqrt{r}}\,dt^2 - \frac{1}{\sqrt{r}\,f(r)}\,dr^2 -
R^2\,\sqrt{r}\,(d\theta_1^2 + \sin^2\theta_1\,d\varphi^2_1) -
\sqrt{r}\,(dx^2 + dy^2)\right)\nn\\
& & + \frac{r^{3/8}}{g^2}\,\left[(\s^1)^2 + (\s^2)^2 + \left(\s^3 +
\frac{g\,Q_e}{\sqrt{2}\,r}\,dt +
\frac{g\,R\,Q_m}{2}\,\cos\theta_1\,d\varphi_1\right)^2\right] +
r^{-5/8}\,dZ^2,\nn\\
\hat{\f} &=& -\frac{1}{4}\,\log(r),\nn\\
\hat{F}_4 &=& -\fr1{g^2}\,\s^2\wedge \s^2\wedge h^3\wedge dZ +
\frac{1}{\sqrt{2}\,g}\,F_2\wedge h^3\wedge dZ,\nn\\
h^3 &=& \s^3 + \frac{g\,Q_e}{\sqrt{2}\,r}\,dt +
\frac{g R\,Q_m}{2}\,\cos\theta_1\,d\varphi_1 \,\,\, .
\ea
For different values of the constants $Q_e, Q_m, R, g$ we will have either a
horizon or a naked singularity.
It should be instructive to compute the entropy  and the Hawking
temperature of this black holes and comparing with the M-theory black
holes obtained in \cite{Klemm:1999in}. Since the system analyzed in that reference is
the same in string variables as the one we analyze here, it is expected
that those results will be repeated. Indeed, the 
ten-dimensional interpretation suggested
in \cite{Klemm:1999in} agrees with the one we have described above.


\noindent
{\em A solution with excited B-fields}

~

Here we consider the six-dimensional Romans' theory with the mass 
parameter and all the fields except the scalar and the three-form field set to zero.  In addition,
we take  $G_3$ to be a constant.
The geometry of the $AdS_3 \times R^3$ spacetime is given by
\beq
ds^2= \rme^{2\,f}\,(dt^2 - dr^2 - dz^2) - \rme^{2\,h}\,(dx^2 + dy^2 + dv^2) \,\,\, .
\eeq
Let us consider a spinor satisfying the following constraints
\beq
\gamma_2 \gamma_7\epsilon_i= \epsilon_i\;\;\; \;\;\;
\gamma_{4 5 6}\epsilon_i=\epsilon_i \,\,\, ,
\eeq
and
\beq
G_{xyv}= G \,\,\, .
\eeq
From the vanishing of the supersymmetric variation of gravitinos and gauginos
we obtain two independent equations
\ba
h^\prime &=& -  \rme^{f}\, \left( \frac{g}{4 \sqrt{2}}\, \rme^\varphi +  \frac{G}{2}\,  \rme^{-3 h + 2\varphi} \right) 
\,\,\, ,\\
f^\prime &=& -  \rme^{f}\, \left( \frac{g}{4 \sqrt{2}}\, \rme^\varphi -  \frac{G}{2} \,  \rme^{-3 h + 2\varphi} \right)\,\,\, ,\\
\varphi^\prime &=&   - h^\prime\,\,\, .
\label{edels}
\ea
A fixed-point solution can be easily obtained
\beq
 \rme^{\frac{\phi}{\sqrt{2}}}=  \frac{g\,\rme^{3h}}{2\sqrt{2} G} \,\,\, .
\eeq
The values of $h(r)$ remain undetermined since 
the equations for $\phi^\prime$ and $h^\prime$
are proportional to each other. In this case, $f(r)$ is given by
\beq
f(r) = - \log \left( \frac{ g^2 \, \rme^{3h} \,
       r}{8 \, G} \right) \,\,\, ,
\eeq
therefore the metric is given by 
\beq
 ds^2 = \frac{1}{\left( \frac{ g^2 \, \rme^{3h} \,
 r}{8 \, G} \right)^2}
 (dt^2 - dr^2 - dz^2)- \rme^{2h} (dx^2 + dy^2 + dv^2) \,\,\, .
\eeq
Equation (\ref{edels}) implies that $\varphi = -h$, where for simplicity 
we omit integration constants. 
The system of equations above can be solved leading to
\beq
f(h)=  - h + \frac{1}{2} \, \log \left( \frac{g}{4 \sqrt{2}} \, \rme^{4h}+\frac{G}{2} \right) \,\,\, .
\eeq
Next, if we make the change of the integration variable $r \rightarrow h$, such that 
\beq
\rme^f\, dr = - \frac{\rme^{5h }}{(\frac{g}{4 \sqrt{2}}\,\rme^{4h})+\frac{G}{2}}\, dh \,\,\, ,
\eeq
in terms of $h$ the metric reads
\beq
 ds^2 =- \frac{\rme^{10h}}{\left(\frac{g}{4 \sqrt{2}}\,\rme^{4h}+ \frac{G}{2}\right)^2}
\,dh^2 + \rme^{-2h}\, \left(
\frac{g}{4 \sqrt{2}}\,
\rme^{4h} + \frac{G}{2}\right)\,(dt^2 - dz^2)- \rme^{2h}\,(dx^2 + dy^2 + dv^2) .
\eeq
Using Eq.(\ref{dualization}) we can
write down the field strength of the dual theory as follows
\be
F_{trz} = -G\,\rme^{2\sqrt{2}\f + 3\,f-3\,h} \,\,\, .
\ee
The ten-dimensional solution is
\ba
ds_{10}^2 &=& -\frac{1}{2}\rme^{\sqrt{2}\f/4}ds_6^2 +
\fr1{g^2}\,\rme^{-3\sqrt{2}\f/4}\,(d\theta^2 + \sin\theta^2 d\varphi^2
+ (d\psi + \cos\theta d\varphi)^2) + \rme^{5\sqrt{2}\f/4}\,dz_1^2,\nn\\
F_4 &=& -G\,\rme^{2\sqrt{2}\f + 3\,f-3\,h}\,dt\wedge
dr\wedge dz\wedge dz_1 - \fr1{g^2}\,\sin\theta\,d\psi\wedge
d\theta\wedge d\varphi\wedge dz_1,\nn\\
\hat{\f} &=& \fr1{\sqrt{2}}\f \,\,\, .
\ea
The metric has a form of a warped product of a six-dimensional space, a three
sphere and a single coordinate.
We also note that the $G_3$ field does not play
a similar role to the previous cases. In fact, there is no twisting in
this case, since the space is not curved. One can view this solution
as a NS-fivebrane of type IIB after wrapping one of the directions on a
circle. It would be interesting to find an example in which the twisting
on the curved manifold is performed by a $B_2$ field.

~

\section{Discussion}

~
Here we would like to start with some comments about the solutions
presented in sections 3 and 4. The
solutions in section 3 are very similar to those obtained in
\cite{Acharya:2000mu} for supergravities in seven dimensions.
This may not be a coincidence since in the string frame both
supergravities (one in seven dimensions and the other in six dimensions)
have very similar structure. Furthermore, their
supersymmetric transformations also look similar. This
indicates that lower-dimensional supergravities are in some sense
related to each other. This relation may be better understood in the
light of the up-lifted solutions in ten or eleven dimensions since the 
up-lifted solutions could
be related by the known dualities.

In section 4 we constructed a non-Abelian solution that, as we noted,
is identical to the solution obtained in 
\cite{Chamseddine:1997nm,Chamseddine:1998mc} and interpreted in
\cite{Maldacena:2001yy} as a wrapped NS-fivebrane. In that case it was a
gravity dual of a theory that is very similar to ${\cal{N}}=1$ SYM. In our 
case, we can
interpret the massless solution as the
same NS-fivebrane with one direction compactified. Therefore, the number
of supercharges remains the same and the world-volume theory is 
three-dimensional ${\cal{N}}=2$ 
SYM theory, as pointed out before. The identity of the string frame action 
in six dimensions (with $m=0$) and the action in seven 
dimensions (with topological mass equal to zero) reveals that
we are dealing with the same system. We can extend this and try to look at
the same issue in five dimensions. In fact, a similar situation occurs 
there too. One of the solutions presented in
\cite{Chamseddine:2001hk} can be related to the NS-fivebrane solution. 
In that paper,
another solution with electric fields is shown. It would be interesting
to understand this electromagnetic solution in the context of string
theory. 

We have analyzed the singularity structure of the equations representing 
D4-branes (of a D4-D8 system) wrapping two- and three-cycles. Because it 
was not possible for us to find exact solutions we turned to approximations
inspired by ref. \cite{Acharya:2000mu}. Thus, we could find some solutions in
Laurent series. We argued that they were non-acceptable as the IR limit of 
a gauge theory. Expanding the solutions around $r=0$ we have identified
the twisted theories on the brane-worldvolume and also found the
operator inserted in the UV of the gauge theory, deforming it and
breaking some supersymmetries. It is desirable to determine the  
precise form of the three (two) dimensional effective action 
of the SCFT that results after wrapping the brane on the two- and three-cycles.

It would be very nice to find new solutions where more fields
are excited. For example solutions with the Abelian gauge field
turned on could be interesting since it seems that they will
involve a non-trivial $B_2$ field. It should also be useful to
find solutions where the twisting is realized by a $B_p$ potential form
(with $p>1$). The relevance of this type of solutions for non-commutative
theories is evident. In this paper we found a solution in  
section 4 that has only $B_2$
fields excited. However, it does not represent a twisted theory since 
the brane is not wrapped
on a curved manifold. Moreover, its ten-dimensional interpretation shows
that we are dealing with a smeared NS-fivebrane.

The work of this paper together with the cases of M5, D3 and M2
branes constitute the cases that complete the
flows between $AdS$-like spacetimes in the UV and the IR limit of gauge 
theories 
in lower dimensions. Finally, it would be  interesting to study systems 
with compactifications like
the ones discussed here but in the context of   $F(4) \times G$
supergravity constructed in 
\cite{D'Auria:2000ad}. For that, it will be necessary to find a prescription
for up-lifting this supergravity to ten dimensions. The
analysis of those theories should give place to theories with enhanced
$E_{Nf +1}$ symmetry, and it should be very interesting to understand the
relation between moduli spaces of Type I' and Heterotic strings from a
gravitational point of view.

~
\newpage
\centerline{\bf Acknowledgements}

We would like to thank Freddy A. Cachazo, Jos\'e Edelstein, Jerome P. Gauntlett, Kentaro Hori, 
Andreas Karch, Nakwoo Kim, Ian I. Kogan,
Juan Maldacena, Christopher N. Pope, Arta Sadrzadeh, Le G.Q. Thong and Cumrun Vafa for
helpful discussions. The work of C.N. is supported by Fundaci\'on
Antorchas of Argentina. The work of I.Y.P. is supported by the US
Department of Energy under grant DE-FG03-95ER40917. 
The work of M.S. is supported in part by funds provided by the U.S. Department of Energy
(D.O.E.) under cooperative research agreement $\#$DF-FC02-94ER40818, CONICET of Argentina, 
Fundaci\'on Antorchas of Argentina and The British
Council. M.S. also acknowledges the kind hospitality of the Theoretical Physics of University 
of Oxford where part of this work was carried out.

~

\vfill

\end{document}